\documentclass{article}

\usepackage{graphicx}
\usepackage{fleqn}
\usepackage{espcrc2}
\usepackage{latexsym}
% ...

\newcommand{\AmS}{{\protect\the\textfont2 
A\kern-.1667em\lower.5ex\hbox{M}\kern-.125emS}} 
\def\etal {{\it et al}. }

\title{
\vspace{-4.0cm} % this length needs adjustment
\begin{flushright}
{\normalsize\tt BNL-HET-99/23}\\
{\normalsize\tt RBRC preprint}\\
\end{flushright}
\vspace*{2.3cm}
The parity partner of the nucleon in quenched QCD with domain 
wall fermions\thanks{Thanks to RIKEN, Brookhaven National Laboratory 
and to the U.S.\ Department of Energy for providing the facilities 
essential for the completion of this work.}}

\author{S.\ Sasaki\address{RIKEN BNL Research Center, 
Brookhaven National Laboratory, NY 11973, USA} (RIKEN-BNL-Columbia 
Collaboration\thanks{The RIKEN-BNL-Columbia Collaboration members 
include T.\ Blum, P.\ Chen, N.\ Christ, M.\ Creutz, C.\ Dawson, G.\ 
Fleming, R.\ Mawhinney, S.\ Ohta, S.\ Sasaki, G.\ Siegert, A.\ Soni, 
P.\ Vranas, M.\ Wingate, L.\ Wu and Y.\ Zhestkov.})}

\begin{document}
%%%%%%%%%%%%%%%%%%%% Abstract %%%%%%%%%%%%%%%%%%%%%%%% 
\begin{abstract}
We present preliminary results for the mass spectrum of the nucleon 
and its low-lying excited states from quenched lattice QCD using 
the domain wall fermion method which preserves the chiral symmetry at 
finite lattice cutoff. Definite mass splitting is observed between 
the nucleon and its parity partner. This splitting grows with 
decreasing valence quark mass. We also present preliminary data 
regarding the first positive-parity excited state.
\end{abstract}

\maketitle

%\section{Introduction}

This work focuses on a notable feature in the mass spectrum of the 
nucleon and its excited states: the mass splitting between the nucleon 
$N(939)$ and its parity partner $N^*(1535)$ is remarkably large 
\cite{PDG}. As is well known, this splitting must be absent if
chiral symmetry were preserved. Yet models with explicit chiral 
symmetry breaking such as non-relativistic quark models or bag models 
fail to reproduce this splitting. In a typical non-relativistic quark 
model with harmonic-oscillator quark wave function \cite{Isgur}, the 
lowest negative-parity state is obtained by adding one oscillator 
quantum to the ground state. The known proton charge radius and 
magnetic moment lead to a few hundred MeV oscillator quantum, far 
underestimating the mass difference. It also gives the wrong ordering of 
positive- and negative-parity excited states: while the 
positive-parity $N'(1440)$ lies below $N^*(1535)$ in nature, the 
model needs two oscillator quanta for $N'$. A similar problem arises 
in bag models where the excitation energy is linked to the inverse of 
the bag radius which in turn is determined by the proton charge radius
\cite{Bag}. 

Thus it is an interesting question whether lattice QCD, 
which appears so successful in describing spontaneous breaking of 
chiral symmetry, can reproduce this mass splitting. Most 
conventional lattice fermion schemes are inadequate for this 
interesting challenge: they break chiral symmetry 
explicitly at finite lattice cutoff \cite{Nielsen} and thus are prone 
to failure in explaining the splitting. 
Fortunately, however, the domain wall fermion (DWF) method seems 
capable of going around this pathology \cite{DWF}. Here we report 
preliminary results of the first quenched calculation of this issue 
using DWF. In this paper, 
24 well separated quenched gauge configurations on a 
$16^3 \times 32$ lattice at 
$6/g^2$=6.0 are used. We use a fifth (DWF) dimension of
$N_s$=16 sites and domain-wall height of $M$=1.8 \cite{RBC}. 

%\section{Baryon operators}

We focus on the spin-half isodoublet baryons. Then there are only two 
possible choices for positive parity baryons 
if we restrict them to contain no derivatives:
$B^{+}_{1}$ =
$\varepsilon_{abc}(u^{T}_{a}C\gamma_{5}d_{b})u_{c}$ and 
$B^{+}_{2}$ = 
$\varepsilon_{abc}(u^{T}_{a}Cd_{b})\gamma_{5}u_{c}$, where
$abc$, $ud$, $C$ and $\gamma_5$ have usual meanings as color, 
flavor, charge conjugation and Dirac matrix. In previous lattice 
calculations of ground-state hadrons, the operator $B^{+}_{1}$ was 
used for the nucleon ground state. Since the operator $B^{+}_{2}$ 
vanishes in the non-relativistic limit, it was considered
ineffective. Indeed, nobody succeeded in extracting the nucleon mass using 
it \cite{Lein}. We will come back to this point later. 

The negative-parity baryon interpolating operators are defined 
with an extra 
$\gamma_{5}$ \cite{FXLee}: $B^{-}_{1}$ = 
$\varepsilon_{abc}(u^{T}_{a}C\gamma_{5}d_{b})\gamma_{5}u_{c}$ and 
$ B^{-}_{2}$ = $\varepsilon_{abc}(u^{T}_{a}Cd_{b})u_{c}$. 
As a result of the definition $B_{1,2}^{-}=\gamma_{5}B_{1,2}^{+}$,
each two-point baryon correlator constructed from 
any one of them actually contains both 
positive- and negative-parity contributions \cite{Fucito}. 
This means that there is contamination from the opposite parity
state propagating backwards in time.
Thus, to extract parity-eigenstate signals
we use a linear combination of quark propagators, one obtained 
with periodic and another with anti-periodic boundary 
conditions in the time direction.

%\section{Results}

We use seven values for the valence quark mass $m$ in the range of 
$0.02 \leq m \leq 0.125$, corresponding to the $\pi\rho$ meson mass 
ratios $m_{\pi}/m_{\rho} \approx 0.59 - 0.90$. Quark propagators 
are calculated with wall source and point sink, and two different 
source positions are used for each gauge configuration. 

%\subsection{Parity partner of nucleon}

Definite plateaus are seen in the effective mass plots for 
$B_{1}^{+}$, $B_{1}^{-}$ and $B_{2}^{-}$ operators. 
%
% FIG 1
%
\begin{figure}
\label{fig:n*}
\begin{center}
\includegraphics[width=72mm]{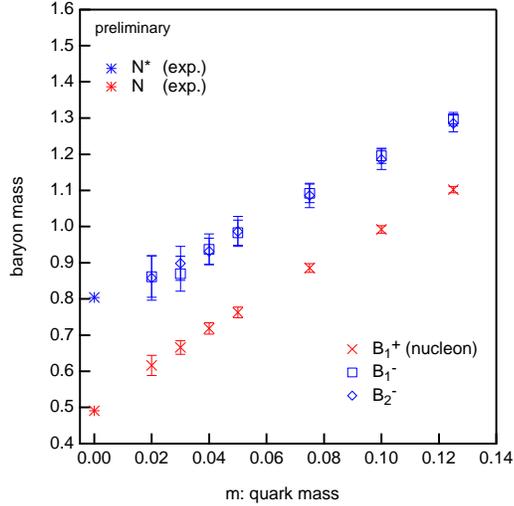} \caption{$N$ ($\times$) 
and $N^*$ ($\Box$ and $\Diamond$) masses versus the quark mass. The 
corresponding experimental values for $N$ and $N^*$ are marked with 
stars. The $N$-$N^*$ mass splitting is clearly observed.}
\end{center}
\end{figure}
In Figure 1, %\ref{fig:n*},
we present our estimates of the nucleon ($N$) and its parity partner 
($N^*$) mass values obtained by taking a weighted average of the 
effective mass in appropriate time ranges. The nucleon mass is 
extracted from the $B_{1}^{+}$ operator. $N^*$ mass estimates from 
$B_{1}^{-}$ and $B_{2}^{-}$ operators agree within errors in the 
whole quark mass range, as expected from their common 
quantum numbers. An important feature is that the $N$-$N^*$ mass 
splitting is observed in the whole range and even for light valence 
quark mass values. Another is that the splitting grows as the valence 
quark mass decreases, suggesting that the large splitting observed in 
nature indeed comes from the spontaneous breaking of chiral symmetry. 
Linear extrapolation in valence quark mass gives us $m_N$=0.56(2) 
and $m_{N^*}$=0.77(2) in lattice units for values in the chiral limit
which are consistent with the experimental value
($a^{-1}\simeq 1.9$ GeV from the $\rho$-meson mass \cite{RBC}).
%
% FIG 2
%
\begin{figure}
\label{fig:ratio}
\begin{center}
\includegraphics[width=72mm]{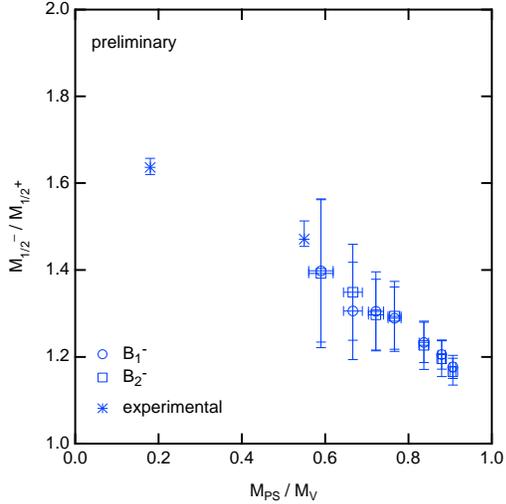} \caption{Mass ratio of 
the negative-parity excited-state and positive-parity ground-state 
baryons versus mass ratio of the pseudoscalar meson and vector meson.}
\end{center}
\end{figure}
In Figure 2, %\ref{fig:ratio},
we compare two mass ratios, one from the baryon parity partners
$m_{N^*}/m_N$ and the other from pseudo-scalar and vector mesons 
$m_\pi/m_\rho$. Experimental points are marked with stars, 
corresponding to non-strange (left) and strange (right) sectors. In 
the strange sector we use $\Sigma$ and $\Sigma(1750)$ as baryon 
parity partners and $K$ and $K^*$ for mesons \cite{PDG}. We find 
the baryon mass ratio grows with decreasing meson mass ratio, toward 
reproducing the experimental values. 
%{\it What do we conclude from this?}

%\subsection{\bf Positive parity signals} 

In contrast to our naive expectation that the operators $B^+_1$ and 
$B^+_2$ should give the same mass estimate, we find different 
plateaus in effective mass plots from these two operators. 
%
% FIG 3
%
\begin{figure}
\label{fig:n'}
\begin{center}
\includegraphics[width=72mm]{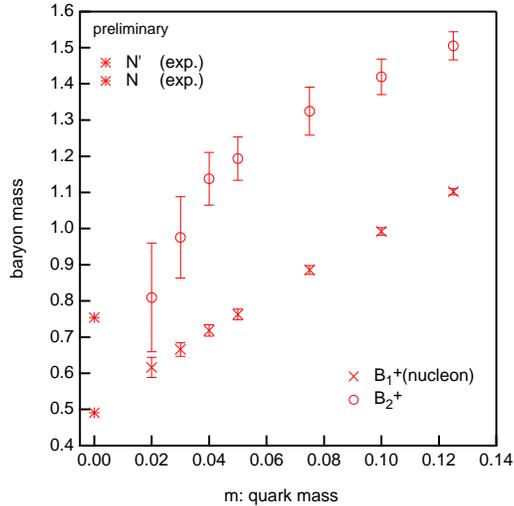} \caption{Mass estimates 
obtained from $B^+_1$ ($\times$) and $B^+_2$ ($\circ$). The 
experimental values for $N$ and $N'$ are marked with stars.}
\end{center}
\end{figure}
In Figure 3, %\ref{fig:n'},
shows that two masses extracted from $B_{1}^{+}$ and $B_{2}^{+}$ are 
quite different.
For heavy quarks ($m \geq 0.04$), we identify $B_{2}^{+}$ with the 
first positive-parity excited state of nucleon ($N'$) for the 
following reasons:
The operator $B_{2}^{+}$ is expected to couple weakly to the ground state 
of the nucleon as we mentioned earlier \cite{Lein}. We suspect the reason 
why we see a clear $B^+_2$ signal for the first time in this study 
while previous studies failed to do so is related to mixing 
induced by explicit chiral symmetry breaking
at finite lattice cutoff which is absent in the former but severe in 
the latter. Although
$B_{1}^{+}$ and
$B_{2}^{+}$ do not mix in the continuum because of different chiral 
structures, it is known that unwanted mixing between them comes 
about through the breaking of chiral symmetry by conventional lattice 
fermions \cite{Richards}.
On the other hand, the DWF exponentially suppresses this breaking and 
thus significantly reduce the unwanted mixing \cite{Aoki}. As a 
result, we are able to numerically confirm an expected feature of 
$\langle 0|B_{2}^{+}|N\rangle\simeq 0$
 at a valence quark mass of $m$=0.04. So for this 
valence quark mass we believe the $B_{2}^{+}$ operator gives an $N'$ 
mass signal. For heavier quark 
mass values, the mass splitting between $N'$ and $N^*$ approaches 
the splitting between $N^*$ and $N$ just like in the naive quark 
or bag models. Unfortunately, however, we have yet to perform the 
$\langle 0|B_{2}^{+}|N \rangle$ calculation for lighter quark mass 
values and hence have not ruled out the possibility that $B_{2}^{+}$ 
couples to the ground-state nucleon. 

%\section{Conclusions}

In conclusion, we have studied the spectrum of the nucleon and its 
excited states by using DWF.
We found the large mass splitting between $N$ and $N^*$ for light 
quark by using two distinct interpolating operators. Our 
$N^*$ mass $m_{N^*}$=0.77(2) in the chiral limit is closer to 
the experimental value than any other study using other fermion 
schemes \cite{FXLee}. We also observed that the unconventional 
nucleon operator gives a clear signal for the first excited nucleon,
at least for heavy quarks.

%%%%%%%%%%%%%%%%%%%%%%%%%%%%

%%%%%%%%%%%%%%%%%%%%%%%%%%%%
\end{document}